\documentstyle[aps,preprint]{revtex}
\begin{document}
\draft
\preprint{\vbox{ \hbox{SOGANG-HEP 237/98} \hbox{SNUTP 98-060} }}
\title{Remarks on self-dual formulation of Born-Infeld-Chern-Simons theory }
\author{Won Tae Kim$^1$\footnote{electronic address:wtkim@ccs.sogang.ac.kr},
  Hyeonjoon Shin$^2$\footnote{electronic address:hshin@kiasph.kaist.ac.kr},
  and Myung Seok Yoon$^1$\footnote{electronic address:younms@physics.sogang.ac.kr}}
\address{$^1$Department of Physics and Basic Science Research Institute, \\ 
        Sogang University, Seoul 121-742, Korea }
\address{$^2$School of Physics, Korea Institute of Advanced Study, 
Seoul 130-012, Korea} 
\date{June 1998}
\maketitle
\bigskip
\begin{abstract}
We study the self-duality of Born-Infeld-Chern-Simons theory which can
be interpreted as a massive D2 brane in IIA string theory
and exhibit the self-dual formulation in terms of  
the gauge invariant master
Lagrangian. The proposed  master Lagrangian contains the nonlocal auxiliary
field and approaches self-dual formulation of Maxwell-Chern-Simons
theory in a point-particle limit with the weak string-coupling limit. 
The consistent canonical brackets of dual system
are derived.  
\end{abstract}

\bigskip

\newpage

Recently, the Born-Infeld(BI) theory is revived through the discovery of
D-branes in the nonperturbative string theory \cite{pol}. The N D-branes are
effectively described by the world volume gauge theory whose action 
is implemented by the
dimensionally reduced U(N) super Yang-Mills action. For a single
D-brane, the effective theory is given by the Dirac-Born-Infeld(DBI) 
action. The massless spectrum of type IIA superstring theory contains 
RR gauge fields apart from NSNS fields. As a RR charge carrier the
D-brane is necessarily introduced. The D2 and D4 are related to the
M2 and M5 in the M-theory \cite{tow95,tow96,sch}. The dual 
description of D-p brane
actions $(p=1,2,3,4)$ are also obtained in Ref. \cite{tse,apps}.
Furthermore the massive
D2 is described by the topological mass term of world volume gauge field 
so called Chern-Simons(CS) term \cite{br,ght,bt} which is connected 
with the 11-dimensional massive supergravity
in terms of the duality transformation \cite{loz}.

On the other hand, the Born-Infeld-Chern-Simon(BICS) theory is 
of interest in its own right apart from the string and M-theory 
sides since the Maxwell-Chern-Simons(MCS) \cite{dj} gauge theory exists in
a certain limiting case. In this paper, we present the master
Lagrangian to dualize the BICS theory. The dual of vector field is not
a scalar but a vector field \cite{loz} and the nonpolynomial BI type
is rewritten by introducing an auxiliary scalar field. Then for the
liming case of string scale and string coupling, the self-duality
of MCS theory can be recovered. Finally we discuss the consistent
brackets for the second class constraints in the dual system of BICS
theory. 

We start with the following master Lagrangian which
is somewhat different in that the standard Born-Infeld action
does not appear apparently,
\begin{equation}\label{master}
{\cal L}_{\rm M}=\tau
      \left[1-\frac{\phi}{2}-\frac{1}{2\phi}\left(1-\kappa^2
      B_nB^n \right)\right] -\tau \kappa^2 \epsilon^{mnp}B_m \partial_n A_p
      +\frac{m}{2}  \epsilon^{mnp} A_m \partial_n A_p ,
\end{equation} 
and the coupling constants are defined by
\begin{equation}
  \tau=\frac{T_2}{g_s}, \qquad \kappa=2\pi {l_s}^2 ,
\end{equation}
where the tension of D2 is $T_2=\frac{1}{4\pi^2l_s^3}$ and $g_s$ is
a string coupling constant and $m$ is a Chern-Simons coefficient. Our
flat metric signature is $\eta_{mn} = {\rm diag}(+1,-1,-1)$. The
master Lagrangian involves the scalar potential 
\begin{equation}
V(\phi)=-1+\frac{\phi}{2}+\frac{1}{2\phi},
\end{equation}
where $\phi$ is an auxiliary scalar field defined by positive value
since the negative one gives a unbound potential.
Thus we take $\phi >0$ and there exists unique stable vacuum $V=0$
for $<\phi>=1$.

It would be interesting to note that the master Lagrangian of
self-dual gauge theory \cite{dj} arises in the the vacuum state
$<\phi>=1$ with $\tau \kappa^2=\frac{1}{e^2}$, 
\begin{equation}\label{sd}
{\cal L}_{\rm SD}=\frac{1}{2}b_n b^n -\epsilon^{mnp}b_m \partial_n a_p
          +\frac{\alpha}{2} \epsilon^{mnp}a_m \partial_n a_p ,
\end{equation}    
after appropriate rescaling, $\frac{1}{e}B_n=b_n$,
$\frac{1}{e}A_n=a_n$, $e^2 m=\alpha$. Therefore the self-dual model 
Eq.(\ref{sd}) from the master Lagrangian Eq.(\ref{master}) corresponds 
to the excitations on the ground state of the potential.

Let us now study the effective theory from the master Lagrangian
(\ref{master}) by classically integrating out 
the auxiliary field and gauge fields.
To obtain the BICS theory, one uses following
equations of motion with respect to $B_n$ and $\phi$, 
\begin{equation}
  \label{B-vari.}
  B^m = \frac12 \phi \epsilon^{mnp} F_{np} ,
\end{equation}
and
\begin{equation}
  \label{phi-vari.}
  \phi = \sqrt{1-\kappa^2 B_m B^m}.
\end{equation}
After eliminating $B_n$ and $\phi$,
the resulting theory becomes, as expected,
\begin{equation}
  \label{bi-cs}
  {\cal L}_{\rm BICS} = \tau \left[ \sqrt{\det \eta_{mn}} -
  \sqrt{\det \left(\eta_{mn} + \kappa F_{mn}\right)} \right] +
  \frac{m}{2} \epsilon^{mnp} A_m \partial_n A_p,
\end{equation}
where we used the identity $\det(G_{mn}+ \kappa F_{mn}) = (\det G_{mn})
\left(1+\frac{\kappa^2}{2} F^{mn}F_{mn} \right)$ for a general metric
$G_{mn}$ in three dimensions where $F_{mn}$ is antisymmetric. 
Note that the BICS theory is approximately reduced to the
Maxwell-Chern-Simons theory,
\begin{equation}
  \label{m-cs}
  {\cal L}_{\rm BICS} \approx -\frac{\tau\kappa^2}{4} F_{mn} F^{mn} +
  \frac{m}{2} \epsilon^{mnp} A_m \partial_n A_p + \tau\kappa^2 O(\kappa^2),
\end{equation}
on the assumption that the string scale and string coupling approach as 
\begin{equation}\label{approx}
l_s \rightarrow 0, \qquad g_s \rightarrow 0. 
\end{equation}
Then $\tau \kappa^2=\frac{l_s}{g_s}$ is fixed and the
higher order of $\kappa^2$ terms are neglected. In some cases
the dynamics of D2 brane may be described by that of
Maxwell-Chern-Simons
theory. The string attached on
the brane in this limit is pointlike and the D2 brane looks like
a massive vector gauge theory. 
On the other hand,  consistent brackets of this theory 
is simply usual Poisson bracket for each variables since the theory
is gauge invariant. 

To obtain the dual description of the BICS theory from 
the master Lagrangian (\ref{master}), the following equation of motion
with respect to $A_m$ 
\begin{equation}\label{A-vari}
  \epsilon^{mnp} \partial_n A_p = \frac{\tau\kappa^2}{m}\epsilon^{mnp}
  \partial_n B_p
\end{equation}
is used. By eliminating $\phi$ and $A_n$
by using Eqs.(\ref{phi-vari.}) and (\ref{A-vari}),
we obtain the dual Lagrangian as
\begin{equation}
  \label{dual}
  {\cal L}_{\rm Dual} = \tau \left[ \sqrt{\det \eta_{mn}} -
  \sqrt{\det \left(\eta_{mn} - \kappa^2 B_m B_n \right)} \right] -
  \frac{(\tau\kappa^2)^2}{2m} \epsilon^{mnp} B_m \partial_n B_p,
\end{equation}
where the relation $\det(G_{mn} - \kappa^2 B_m B_n) = (\det G_{mn}) (1-
\kappa^2 B_m B^m)$ for odd dimension is used.
In this dual relation between the BICS and dual version, the
duality has some significance on the interactions.
If one want to explain the BICS theory in terms of usual field theory
the higher derivatives are infinitely generated in the expended form 
while the maximum order of derivative is linear in the dual system.  
Then the higher order of momentum transfer 
in the vertices is understood 
as the infinitely many contact interactions in the dual version.
So the high momentum interaction may be regarded as the low momentum
interaction.

On the other hand, 
the dual system is gauge-noninvariant and has a symplectic structure
due to the Chern-Simons term. Similarly to the BICS case, this dual
Lagrangian can be approximately reduced to
\begin{equation}\label{dualm-cs}
  {\cal L}_{\rm Dual} \approx \frac{\tau\kappa^2}{2} B_m B^m -
  \frac{(\tau\kappa^2)^2}{2m} \epsilon^{mnp} B_m \partial_n B_p +
  \tau\kappa^2 O(\kappa^2  )
\end{equation}
for the limit (\ref{approx}). The fact that Eqs.(\ref{m-cs})and
(\ref{dualm-cs}) are dual is already shown through the self-dual
master Lagrangian (\ref{sd}) in Ref. \cite{dj}. 
 
In the dual description of the BICS theory of Eq.(\ref{dual}) 
all constraints will belong to second class, 
so consistent brackets are nontrivial.
To obtain consistent brackets in the dual theory, we
obtain primary constraints   
from Eq.(\ref{dual}), 
\begin{eqnarray}
  \Omega_1 &=& \Pi_B^0 \approx 0, \\
  \Omega_2 &=& \Pi_B^1 + \frac{(\tau\kappa^2)^2}{2m}B_2 \approx 0, \\
  \Omega_3 &=& \Pi_B^2 - \frac{(\tau\kappa^2)^2}{2m}B_1 \approx 0,
\end{eqnarray}
and Hamiltonian
\begin{equation}
  H_p = \int\/ d^2 x \left\{ \tau\left[ \sqrt{1-\kappa^2B_m B^m} -
  1\right] - 2 B_0 \partial_i \Pi_B^i + \lambda_1 \Omega_1 +
  \lambda_2 \Omega_2 + \lambda_3 \Omega_3\right\},
\end{equation}
with the Gauss constraint,
\begin{equation}
  \Omega_4 =\dot{\Omega}_1= 2\partial_i \Pi_B^i - \frac{\tau\kappa^2
  B_0}{\sqrt{1-\kappa^2 B_m B^m}} \approx 0.
\end{equation}
The Gauss' law constraint is given by the stability condition
of the primary constraint $\Omega_1$ and the three Lagrange multiplier 
fields are completely fixed through the time stability of constraints.
Therefore we can define a Dirac matrix $C_{ab}(x,y) = \{ \Omega_a
(x),\Omega_b(y)\}$, and obtain the inverse Dirac matrix 
$C_{ab}^{-1}(x,y)$ \cite{dir}. The Dirac Brackets are defined by
\begin{eqnarray}
  \{A(x),B(y)\}_{\rm D} &=& \{A(x),B(y)\}
  \nonumber \\
  & &\ \  - \sum_{a,b} \int\int d^2 z d^2 w \{ A(x),\Omega_a
  (z)\} C_{ab}^{-1}(z,w) \{ \Omega_b(w), B(y)\},
\end{eqnarray}
where $a,b=1,\cdots,4$. According to this
definition,
we obtain the nonvanishing brackets, 
\begin{eqnarray}
\{ B_0 (x) , B_0 (y)\}_{\rm D} &=& \partial_i \left[ 
  \frac{B_0 B^i (1-\kappa^2 B_m B^m )^{3/2}}{\tau\kappa^2
  (1+\kappa^2 {\bf B}^2)^2}\right]\delta^2({\bf x} - {\bf y})\label{1}
  \nonumber \\
  & & \qquad \qquad + \ \left[ \frac{2 B_0 B^i (1-\kappa^2 B_m B^m
  )^{3/2}}{\tau\kappa^2(1+\kappa^2 {\bf B}^2)^2}\right]\partial_i
  \delta^2({\bf x} - {\bf y}) , \\
\{ B_0(x), B^i(y)\}_{\rm D} &=&
  -\frac{m}{(\tau\kappa^2)^2} \frac{\epsilon^{ij}B_0
  B_j}{1+\kappa^2 {\bf B}^2} \delta^2({\bf x} - {\bf y}) , \\
\{ B_0(x),\Pi_B^i(y)\}_{\rm D} &=& - \frac{B_0
  B^i}{2(1+\kappa^2{\bf B}^2)} \delta^2({\bf x}- {\bf y}) +
  \frac{\tau\kappa^2 (1-\kappa^2 B_m
  B^m)^{3/2}}{2m(1+\kappa^2{\bf B}^2)}
  \epsilon^{ij}\partial_j \delta^2({\bf x} - {\bf y}) , \\
\{ B_1 (x), B_2(y)\}_{\rm D} &=& -
  \frac{m}{(\tau\kappa^2)^2} \delta^2({\bf x} - {\bf y}),\label{ind} \\
\{ B_1 (x), \Pi_B^1(y)\}_{\rm D} &=& \frac12
  \delta^2({\bf x} - {\bf y}) ,\\
\{ B_2 (x), \Pi_B^2(y)\}_{\rm D} &=& \frac12
  \delta^2({\bf x} - {\bf y}), \\
\{ \Pi_B^1 (x), \Pi_B^2(y)\}_{\rm D} &=& -
  \frac{(\tau\kappa^2)^2}{4m} \delta^2({\bf x} - {\bf y})\label{2}. 
\end{eqnarray}
The independent bracket is just only Eq.(\ref{ind}) which
is compatible with the previous result in Ref. \cite{dj}.
The other brackets are in fact derived from this fundamental
bracket Eq.(\ref{ind}). Note that the brackets for the
limit (\ref{approx}) are well defined in Eq. (\ref{1})-(\ref{2}).

After using the four constraints,
 the reduced Hamiltonian is written in the form of
\begin{eqnarray}\label{reduced}
  H &=& \int d^2 x \frac{\tau}{\sqrt{1-\kappa^2B_m B^m}}
  \left[ 1+\kappa^2 {\bf B}^2 - \sqrt{1+\kappa^2 {\bf B}^2 -\kappa^2
  (B_0)^2} \right] \nonumber \\
    &=& \int d^2 x \frac{\tau}{\kappa\sqrt{{\bf B}^2}} \left[
  (1+\kappa^2{\bf B}^2) \sqrt{1+\frac{\kappa^2}{m^2}
  (\tau\kappa^2)^2\epsilon^{ij}\partial_i B_j} -
  \kappa\sqrt{{\bf B}^2} \right].
\end{eqnarray}
In the first line of Eq.(\ref{reduced}), we find that the
Hamiltonian is positive definite and in the second line the
Hamiltonian is rewritten in terms of independent degrees of freedom.
  
In summary, we have studied the master Lagrangian to relate the BICS
theory and dual theory which is apparently unrelated. It is a
generalization of dual formulation of the MCS theory. From the point
of D2 brane, the master Lagrangian for the limit $l_s \rightarrow 0$
and $g_s \rightarrow 0 $ reproduces the self dual gauge theory.   
 
{\bf Acknowledgments}\\
%\section*{Acknowledgments}
This work was supported in part by Ministry of Education, 1997, Project
No. BSRI-97-2414, and Korea Science and Engineering Foundation through
the Center for Theoretical Physics in Seoul National University(1998).
%%%%%%%%%%%%%%%%%%%% References %%%%%%%%%%%%%%%%%%%%%%%%%

\end{document}